\documentclass[conference]{IEEEtran}
\IEEEoverridecommandlockouts
\usepackage{cite}
\usepackage{amsmath,amssymb,amsfonts}
\usepackage{algorithmic}
\usepackage{graphicx}
\usepackage{textcomp}
\usepackage{xcolor}

\usepackage{bm,cite}

\newcommand{\minimize}{\mathop{\rm minimize}\limits}

\def\H{{\mathsf{H}}}

\def\del{\partial}

\def\j{{\mathrm j}}

\def\M{{\mathcal M}}

\def\BibTeX{{\rm B\kern-.05em{\sc i\kern-.025em b}\kern-.08em
    T\kern-.1667em\lower.7ex\hbox{E}\kern-.125emX}}
\begin{document}

\title{Multichannel Active Noise Control with\\Exterior Radiation Suppression\\Based on Riemannian Optimization
}

\author{\IEEEauthorblockN{Takaaki Kojima}
\IEEEauthorblockA{\textit{Faculty of Engineering, The University of Tokyo}\\
Tokyo, Japan \\
kojima-takaaki@g.ecc.u-tokyo.ac.jp}
\and
\IEEEauthorblockN{Kazuyuki Arikawa}
\IEEEauthorblockA{\textit{Graduate School of Information Science and Technology,} \\
\textit{The University of Tokyo}\\
Tokyo, Japan}
\and
\IEEEauthorblockN{Shoichi Koyama}
\IEEEauthorblockA{\textit{Digital Content and Media Sciences Research Division,} \\
\textit{National Institute of Informatics}\\
Tokyo, Japan}
\and
\IEEEauthorblockN{Hiroshi Saruwatari}
\IEEEauthorblockA{\textit{Graduate School of Information Science and Technology,} \\
\textit{The University of Tokyo}\\
Tokyo, Japan}
}

\maketitle

\begin{abstract}
A multichannel active noise control (ANC) method with exterior radiation suppression is proposed. When applying ANC in a three-dimensional space by using multiple microphones and loudspeakers, the loudspeaker output can amplify noise outside a region of target positions because most of current ANC methods do not take into consideration the exterior radiation of secondary loudspeakers. We propose a normalized least mean square algorithm for feedforward ANC in the frequency domain based on the Riemannian optimization to update the control filter with the exterior radiation power constrained to a target value. The advantages of the proposed method, compared with the algorithm using a penalty term of exterior radiation, were validated by numerical experiments: the exterior radiation power can be constrained during the adaptation process and the parameter for the constraint can be determined in advance.  
\end{abstract}


\section{Introduction}
\label{sec:intro}


The goal of active noise control (ANC) is to cancel unwanted noise from primary noise sources using secondary loudspeakers. In typical multichannel feedforward ANC systems, the driving signals of the secondary loudspeakers to reduce noise at positions of error microphones, i.e., \textit{target positions}, are obtained from reference microphone signals by filtering through a control filter adaptively optimized on the basis of error microphone signals. Although ANC techniques have been studied for several decades~\cite{nelson1991active,kuo1999active,kajikawa_gan_kuo_2012}, their application to a three-dimensional (3D) space has recently attracted attention again because of recent advancements on spatial ANC techniques~\cite{Samarasinghe_2016_InsideCabin,Zhang_2018_ANCoverSpace,ma2020,Koyama:IEEE_ACM_J_ASLP2021}.

When applying the ANC techniques in a 3D space, an exterior region of target positions for noise reduction is normally not taken into consideration. Therefore, the noise outside the region of target positions can be largely amplified owing to secondary loudspeaker outputs. Several attempts have been made to suppress the output power of secondary loudspeakers~\cite{rafaely2000computationally, qiu2001study,shi2019two,shi2021opt}; however, the reduction in output power does not always lead to the suppression of exterior radiation. 

For the above reasons, it is important to develop a multichannel ANC method to suppress the exterior radiation power of secondary loudspeakers while reducing noise at the target positions. The exterior radiation power can be formulated with their given directivity patterns~\cite{Ueno:ICASSP2018}. In our previous study~\cite{Arikawa:ICA2022}, normalized least mean square (NLMS)-based adaptive filtering algorithms for feedforward ANC in the frequency domain are derived with a penalty term or an inequality constraint on this exterior radiation formulation in the context of spatial ANC. To adapt the constraint on the external radiation power to the primary noise amplitude with the alleviation of the effect on the ANC performance, the NLMS algorithm derived from the cost function with an additive penalty term for exterior radiation power can be used. However, this algorithm has several issues in practice: 1) the exterior radiation power is not necessarily suppressed during the adaptation process even when its target value is successfully reached after convergence and 2) it is difficult to determine the parameter for balancing the penalty term before the adaption process. 

We propose an NLMS-based multichannel feedforward ANC algorithm in the frequency domain based on the Riemannian optimization~\cite{AbsMahSep2008, sato2021riemannian}, which can be regarded as an application of stochastic gradient descent on Riemannian manifolds~\cite{stochastic}. The optimization problem for computing the control filter is defined as the minimization problem of noise at the target positions with an equality constraint on the exterior radiation power. We developed an NLMS algorithm to update the control filter on a Riemannian manifold constructed by the equality constraint. We conducted numerical experiments to evaluate the performance of the proposed method.






\section{Multichannel ANC}%
\label{sec:ANC}

\begin{figure}[tb]
  \begin{minipage}[b]{1.0\linewidth}
    \centering
    \centerline{\includegraphics[width=7.0cm]{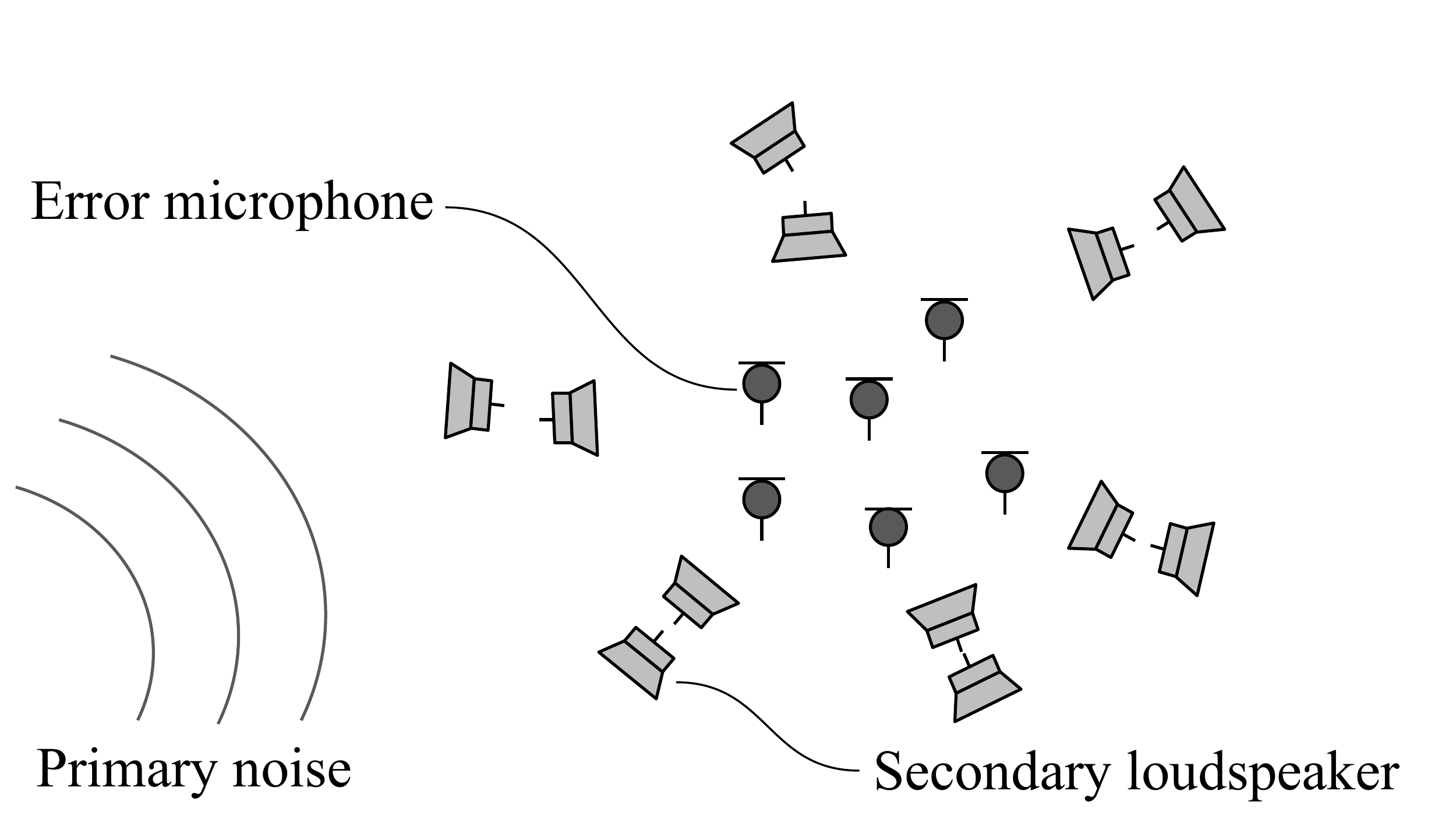}}
  \end{minipage}
  \caption{Multichannel ANC in 3D space using multiple microphones and loudspeakers.}
  \label{fig:ANC_arrangement}
  \vspace{-0.5cm}
\end{figure}

Suppose that $L$ secondary loudspeakers and $M$ error microphones are placed in a 2D or 3D acoustic space, i.e., $\mathbb{R}^2$ or $\mathbb{R}^3$, as shown in Fig.~\ref{fig:ANC_arrangement}. $R$ reference microphones are placed near primary noise sources. The observed signals of the error and reference microphones, and the driving signals of the secondary loudspeakers at time frame $n$ and angular frequency $\omega$ are denoted by $\bm{e}_n(\omega)\in\mathbb{C}^M$, $\bm{x}_n(\omega)\in\mathbb{C}^R$, and $\bm{y}_n(\omega)\in\mathbb{C}^L$, respectively. By denoting the primary noise at error microphone positions as $\bm{d}_n(\omega)\in\mathbb{C}^M$, we express the error microphone signals as
\begin{align}
\bm{e}_n(\omega) &= \bm{d}_n(\omega)+\bm{G}(\omega)\bm{y}_n(\omega) \notag\\
&= \bm{d}_n(\omega)+\bm{G}(\omega)\bm{W}_n(\omega)\bm{x}_n(\omega),
\label{eq:error_signal}
\end{align}
where $\bm{G}(\omega)\in\mathbb{C}^{M\times L}$ is the transfer function matrix from secondary loudspeakers to error microphones and $\bm{W}_n(\omega)\in\mathbb{C}^{L\times R}$ is the adaptive control filter to obtain the optimal driving signals from the reference signals. Hereafter, the argument $\omega$ is omitted for notational simplicity. 


The cost function of the multichannel ANC is generally defined as the expectation value of the power of error signals:
\begin{align}
    J=\mathbb{E}[\sigma_{n}(\bm{W}_n)] 
\label{eq:Jint}
\end{align}
with
\begin{align}
    \sigma_{n}(\bm{W}_n) := \|\bm{d}+\bm{G}\bm{W}_n\bm{x}_n\|_2^2+\gamma\|\bm{W}_n\bm{x}_n\|_2^2,
\end{align}
and a regularization parameter $\gamma>0$. By replacing the expectation value of $\sigma_{n}$ with the instantaneous value in \eqref{eq:Jint}, i.e., $J \approx \sigma_{n}(\bm{W}_n)$, the NLMS algorithm for updating $\bm{W}_n$ is derived as
\begin{align}
  \bm{W}_{n+1}&=\bm{W}_n-\mu_n\frac{\del \sigma_{n}}{\del \bm{W}_n^*} \notag\\
  &=\bm{W}_n-\mu_n(\bm{G}^\H\bm{e}_n+\gamma\bm{W}_n\bm{x}_n)\bm{x}_n^\H,
\label{eq:MPC-NLMS} 
\end{align}
with the step size parameter \cite{haykin2013adaptive}
\begin{align}
  \mu_n=\frac{\mu_0}{\|\bm{G}^\H\bm{G}+\gamma \bm{I}_L\|_2\|\bm{x}_n\|_2^2}.
  \label{eq:MPC-NLMS_step} 
\end{align}
Here, $\bm{I}_L$ is the identity matrix of size $L \times L$ and $\mu_0\in(0,2)$ is a normalized step size parameter. $(\cdot)^\ast$ and $(\cdot)^\H$ denote the complex conjugate and conjugate transpose, respectively.

\section{Exterior Radiation Suppression with Penalty Term}
\label{sec:ANC-ExRadSupp}
\vspace{-0.2cm}

We introduce a multichannel ANC method to suppress the exterior radiation power while reducing noise, which is proposed in \cite{Arikawa:ICA2022} in the context of spatial ANC.   


First, the exterior radiation power of the secondary loudspeakers is formulated. Let $\del \Omega$ be a surface of a circular or spherical area including all the secondary loudspeakers. The total acoustic power radiated from $\del \Omega$ by all the secondary loudspeakers, that is, the exterior radiation power, is defined for the pressure field $u_{n}(\bm{r})$ generated by the $n$th-frame driving signals of the secondary loudspeakers as \cite{Ueno:ICASSP2018}
\begin{align}
  \varepsilon_{n}(\bm{W}_n) := \int_{\del \Omega}\frac{1}{2}\mathrm{Re}\left[u_{n}(\bm{r})\frac{\j}{\rho c k}\frac{\del u_{n}(\bm{r})^{\ast}}{\del\bm{n}}\right]\mathrm{d}\bm{r},
  \label{eq:Jext}
\end{align}
where $\rho$ is the medium density, $c$ is the speed of sound, $k:=\omega/c$ is the wave number, and $\del / \del \bm{n}$ denotes the normal derivative on $\del \Omega$. By representing $u_{n}(\bm{r})$ with the $n$th-frame driving signals $\bm{y}_n$, $\varepsilon_{n}(\bm{W}_n)$ is reformulated with a Hermitian matrix $\bm{A}\in\mathbb{C}^{L\times L}$ as 
\begin{align}
  \varepsilon_{n}(\bm{W}_n)=\bm{y}_n^\H\bm{A}\bm{y}_n=\bm{x}_n^\H\bm{W}_n^\H\bm{A} \bm{W}_n\bm{x}_n.
  \label{eq:Jext-2}
\end{align}
When all the secondary loudspeakers are point sources, the $(l,l^\prime)$th element of $\bm{A}$ can be expressed as~\cite{Ueno:ICASSP2018}
\begin{align}
  (\bm{A})_{l,l^\prime}=
  \begin{cases}
    \displaystyle \frac{1}{8c\rho k}J_0(k\|\bm{r}_l-\bm{r}_{l^\prime}\|_2)&\text{in 2D space}\vspace{6pt}\\
    \displaystyle \frac{1}{8c\rho k}j_0(k\|\bm{r}_l-\bm{r}_{l^\prime}\|_2)&\text{in 3D space}
  \end{cases},
  \label{eq:Jext-elements}
\end{align}
where $\bm{r}_l$ is the position of the $l$th secondary loudspeaker, and $J_0(\cdot)$ and $j_0(\cdot)$ are 0th-order Bessel function and 0th-order spherical Bessel function, respectively. Note that $\varepsilon_n$ corresponds to the power of the driving signals of the secondary loudspeakers when $\bm{A}$ is  $\bm{I}_L$.

A simple strategy to suppress the exterior radiation power while reducing noise at the error microphone positions is to define the cost function as the weighted sum of $\sigma_{n}$ and $\varepsilon_{n}$ as follows~\cite{Arikawa:ICA2022}:
\begin{align}
      J_{\mathrm{Penal}} = \sigma_{n}(\bm{W}_n)+\lambda \varepsilon_{n}(\bm{W}_n),
  \label{eq:Jpenalty}
\end{align}
where $\lambda>0$ is the parameter used to determine the balance of the two terms. We here defined $J_{\mathrm{Penal}}$ as the instantaneous value instead of the expectation value as in the NLMS algorithm \eqref{eq:MPC-NLMS}. The NLMS algorithm for minimizing $J_{\mathrm{Penal}}$ is derived similarly as
\begin{align}
\bm{W}_{n+1}=\bm{W}_n-\mu_n[\bm{G}^\H\bm{e}_n+(\gamma\bm{I}_L+\lambda\bm{A})\bm{W}_n\bm{x}_n]\bm{x}_n^\H,
\label{eq:MPC-Penalty-NLMS}
\end{align}
with the step size parameter
\begin{align}
  \mu_n=\frac{\mu_0}{\|\bm{G}^\H\bm{G}+\gamma \bm{I}_L+\lambda\bm{A}\|_2\|\bm{x}_n\|_2^2}.
\end{align}



\section{Proposed Algorithm Based on Riemannian Optimization}
\label{sec:ANC-Equality}

\subsection{Riemannian Optimization for Exterior Radiation Suppression}

The NLMS algorithm with a penalty term of the exterior radiation power presented in Sect.~\ref{sec:ANC-ExRadSupp} has several issues in practice. First, although the exterior radiation can be suppressed after the convergence of the adaptive filter, there is no guarantee that the exterior radiation is suppressed during the adaptation process. Second, it is not simple to determine an appropriate parameter $\lambda$ because it is difficult to explicitly relate the exterior radiation power after convergence with the parameter $\lambda$. 

To overcome the above issues, we propose an NLMS algorithm based on the Riemannian optimization with an equality constraint on the exterior radiation power. We define the optimization problem as
\begin{align}
    &\minimize_{\bm{W} \in \mathbb{C}^{L\times R}} \ \ \sigma_{n}(\bm{W}) \notag\\
    &\mathrm{subject~to} \ \ \bm{W}^\H\bm{A}\bm{W}=C\bm{I}_{R},
\label{eq:MPC-equality-problem}
\end{align}
where $C>0$. Again, the optimization problem is defined with the instantaneous value instead of the expectation value. By using this equality constraint, we can constrain the exterior radiation power proportional to the power of the reference microphones as
\begin{align}
  \varepsilon_{n}(\bm{W}) = \bm{x}_n^\H\bm{W}_n^\H\bm{A}\bm{W}_n\bm{x}_n=C\|\bm{x}_n\|_2^2.
  \label{eq:MPC-Equality-Prad}
\end{align}
We represent the equality constraint in \eqref{eq:MPC-equality-problem} as a Riemannian manifold $\M$ on which the control filter $\bm{W}$ is updated:
\begin{align}
  \M := \{\bm{W}\in\mathbb{C}^{L\times R} \ | \ \bm{W}^\H\tilde{\bm{A}}\bm{W}=\bm{I}_{R}\},
\end{align}
where $\tilde{\bm{A}}:=\bm{A}/C$. Thus, the adaptive algorithm is obtained for the unconstrained minimization problem of $\sigma_{n}(\bm{W})$ on $\M$.




\subsection{Derivation of NLMS algorithm}

The control filter $\bm{W}$ is regarded as a point on a generalized Stiefel manifold $\M$. By defining an appropriate Riemannian metrics for $\M$, we obtain the gradient of the cost function, $\mathrm{grad} \ \sigma_{n}(\bm{W})\in T_{\bm{W}}\M$, as an orthogonal projection of the standard gradient $\del \sigma_{n}(\bm{W})/\del \bm{W}^{\ast}$ onto the tangent space of $\M$ at $\bm{W}$, denoted by $\mathcal{T}_{\bm{W}}\M$~\cite{AbsMahSep2008}. To update $\bm{W}$ in the steepest descent direction, an approximate mapping called  retraction is used~\cite{AbsMahSep2008}.


The orthogonal projection of $\bm{U}\in \mathcal{T}_{\bm{W}}\mathbb{C}^{L\times R}$ onto $\mathcal{T}_{\bm{W}}\M$ is defined as~\cite{Shustin2019_projection}
\begin{align}
  \mathcal{P}_{\bm{W}}(\bm{U})=\bm{U}-\tilde{\bm{A}}\bm{W}\bm{H},
  \label{eq:def-projection}
\end{align}
where $\bm{H}$ is the unique solution of the following Sylvester equation~\cite{horn2012matrix}:
\begin{align}
  \notag(\bm{W}^\H\tilde{\bm{A}}^\H\tilde{\bm{A}}\bm{W})\bm{H}+\bm{H}(\bm{W}^\H\tilde{\bm{A}}^\H\tilde{\bm{A}}\bm{W})\\
  =\bm{W}^{\H}\tilde{\bm{A}}\bm{U}+\bm{U}^{\H}\tilde{\bm{A}}\bm{W}.
\label{eq:Sylvester}
\end{align}
Then, the gradient of the cost function is represented as
\begin{align}
  \mathrm{grad} \ \sigma_{n}(\bm{W}) = \mathcal{P}_{\bm{W}}\left(2(\bm{G}^\H\bm{e}_n+\gamma\bm{W}\bm{x}_n)\bm{x}_n^\H \right).
  \label{eq:riemannian-grad-Jint}
\end{align}


Next, a retraction at $\bm{W}\in\M$ from $\bm{V}\in \mathcal{T}_{\bm{W}}\M$ can be defined as~\cite{Sato2019}
\begin{align}
  \mathcal{R}_{\bm{W}}(\bm{V})=\sqrt{\tilde{\bm{A}}}^{-1}\mathrm{qf}\left(\sqrt{\tilde{\bm{A}}}(\bm{W}+\bm{V})\right),
  \label{eq:retraction}
\end{align}
where $\mathrm{qf}(\cdot)$ denotes the function that returns the Q-factor of QR factorization when all the diagonal elements of the R-factor are positive. Note that $\tilde{\bm{A}}$ as well as $\bm{A}$ is assumed to be positive definite.


The proposed NLMS algorithm is summarized as an iteration of the following steps, starting with an initial value $\bm{W}_0\in \M$:
\begin{enumerate}
    \item Compute the gradient $\mathrm{grad}\ \sigma_{n}(\bm{W}_n)$ in \eqref{eq:riemannian-grad-Jint}.
    \item Update $\bm{W}_{n+1}=\mathcal{R}_{\bm{W}}(-\mu_n \ \mathrm{grad}\ \sigma_{n}(\bm{W}_n))$.
\end{enumerate}
We also update the step size parameter $\mu_n$ on the basis of the power of the reference signal $\|\bm{x}_n\|_2^2$ at each iteration according to \eqref{eq:MPC-NLMS_step} as
\begin{align}
\mu_n =\frac{\mu_0}{2{\|\bm{G}^\H\bm{G}+\gamma \bm{I}_L\|_2\|\bm{x}_n\|_2^2}}.
\end{align}
In the above steps, additional computations after calculating the standard gradient $\del \sigma_n(\bm{W})/\del\bm{W}^{\ast}$, whose computational cost $O(ML+2LR)$ is equivalent to that of the NLMS algorithm in \eqref{eq:MPC-NLMS}, are necessary at each iteration. The gradient $\mathrm{grad}~\sigma_n(\bm{W})$ requires $O(R^3)$ for solving the Sylvester equation \eqref{eq:Sylvester}~\cite{bartels1972solution} and $O(LR^2+L^2R)$ for the orthogonal projection \eqref{eq:def-projection}. The retraction \eqref{eq:retraction} requires $O(LR^2)$ for matrix manipulation and $O(R^3)$ for QR factorization. Since the number of reference microphones $R$ is generally small, the increase in the total computational cost from \eqref{eq:MPC-NLMS} is not large in practice.

\begin{figure}[tb]
  \begin{minipage}[b]{1.0\linewidth}
    \centering
    \centerline{\includegraphics[width=7.4cm]{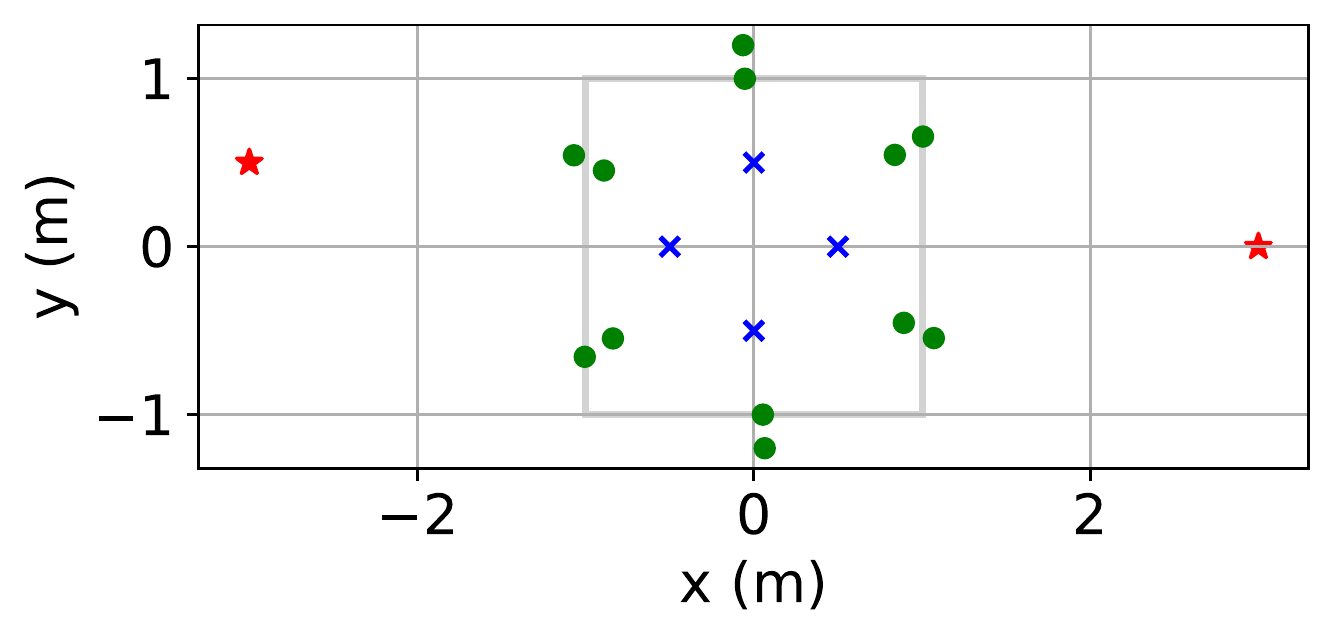}}
  \end{minipage}
  \caption{Experimental settings. Blue crosses, green dots, and red stars indicate error microphones, secondary loudspeakers, and primary noise sources, respectively.}
  \label{fig:setting}
\end{figure}

\section{Numerical Experiments}
\label{sec:experiments}

We conducted numerical experiments to evaluate the performance of the proposed method in terms of noise reduction and exterior radiation suppression  in a 2D free field. We compared the NLMS algorithm without exterior radiation suppression (\textbf{NLMS}), the NLMS algorithm based on the penalty term~\cite{Arikawa:ICA2022} (\textbf{Ext-Penal NLMS}), and the proposed method based on the Riemannian optimization  (\textbf{Ext-Riem NLMS}).

\begin{figure}[tb]
  \begin{minipage}[b]{1.0\linewidth}
    \centering
    \centerline{\includegraphics[width=8.4cm]{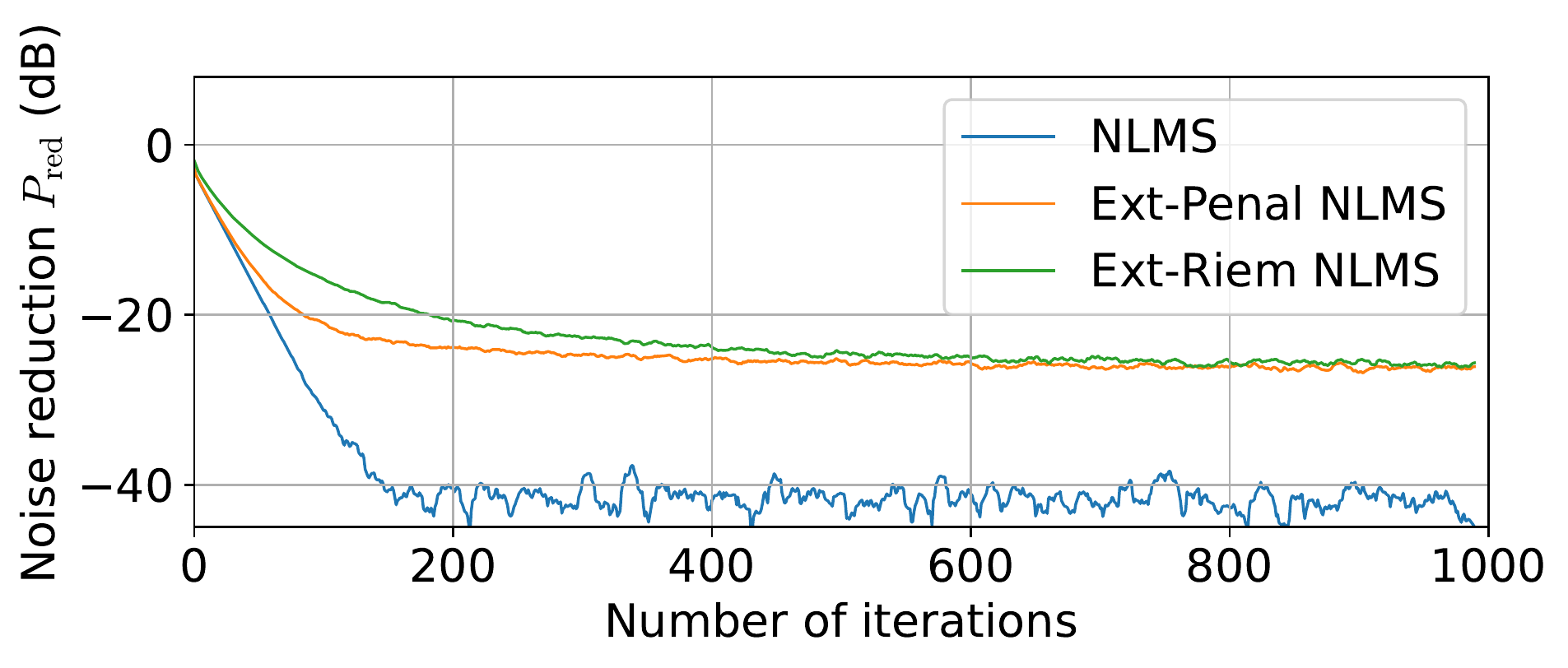}}
    \centerline{(a) Noise reduction $P_{\mathrm{red}}$}
  \end{minipage}
  \begin{minipage}[b]{1.0\linewidth}
    \centering
    \centerline{\includegraphics[width=8.4cm]{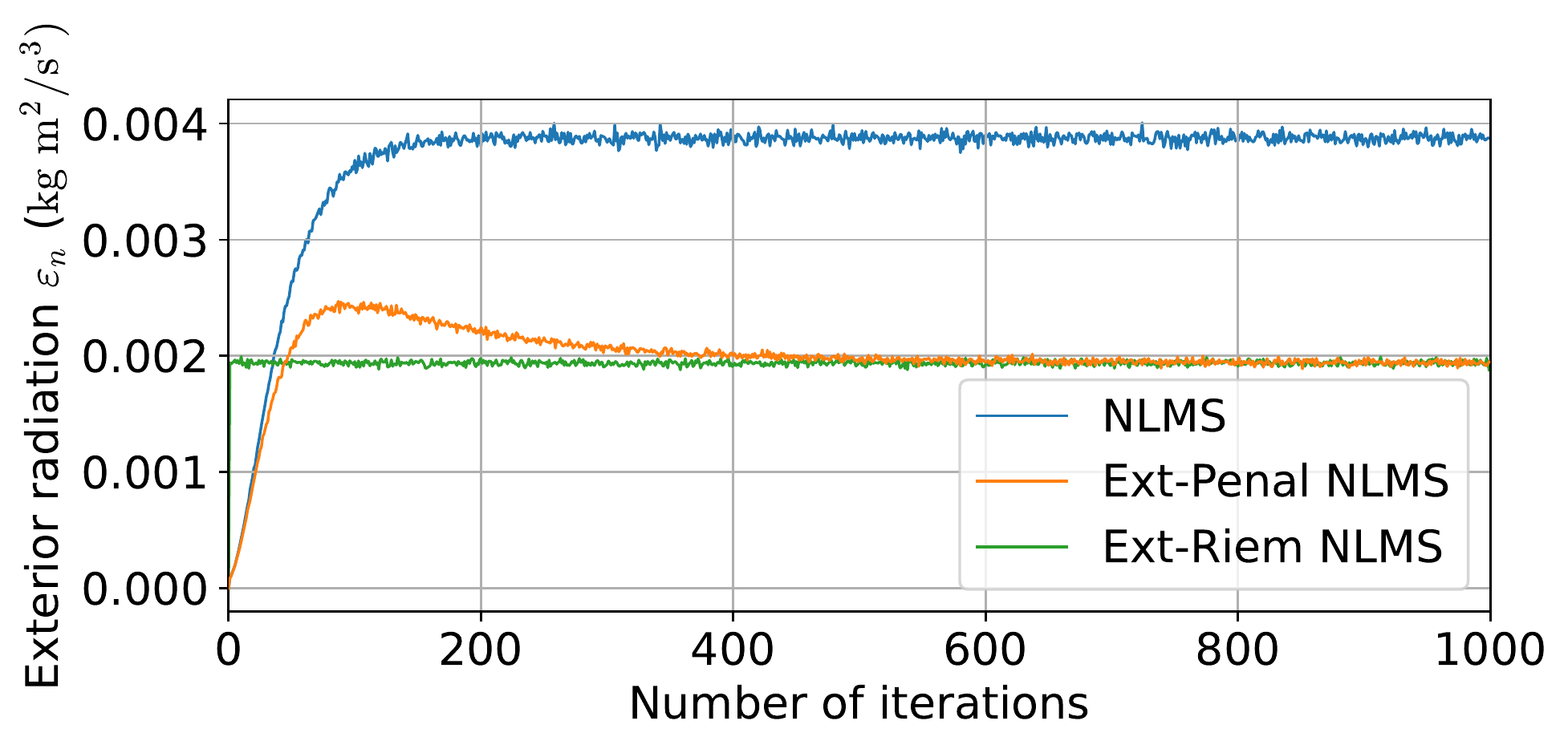}}
    \centerline{(b) Exterior radiation $\varepsilon_n$}
  \end{minipage}%
  \caption{Noise reduction $P_{\mathrm{red}}$ and exterior radiation $\varepsilon_n$ at each iteration when the noise frequency was 500 Hz.}
  \label{fig:result-500Hz-normal}
\end{figure}
\begin{figure}[t]
  \begin{minipage}[b]{\linewidth}
    \centering
    \centerline{\includegraphics[width=8.4cm]{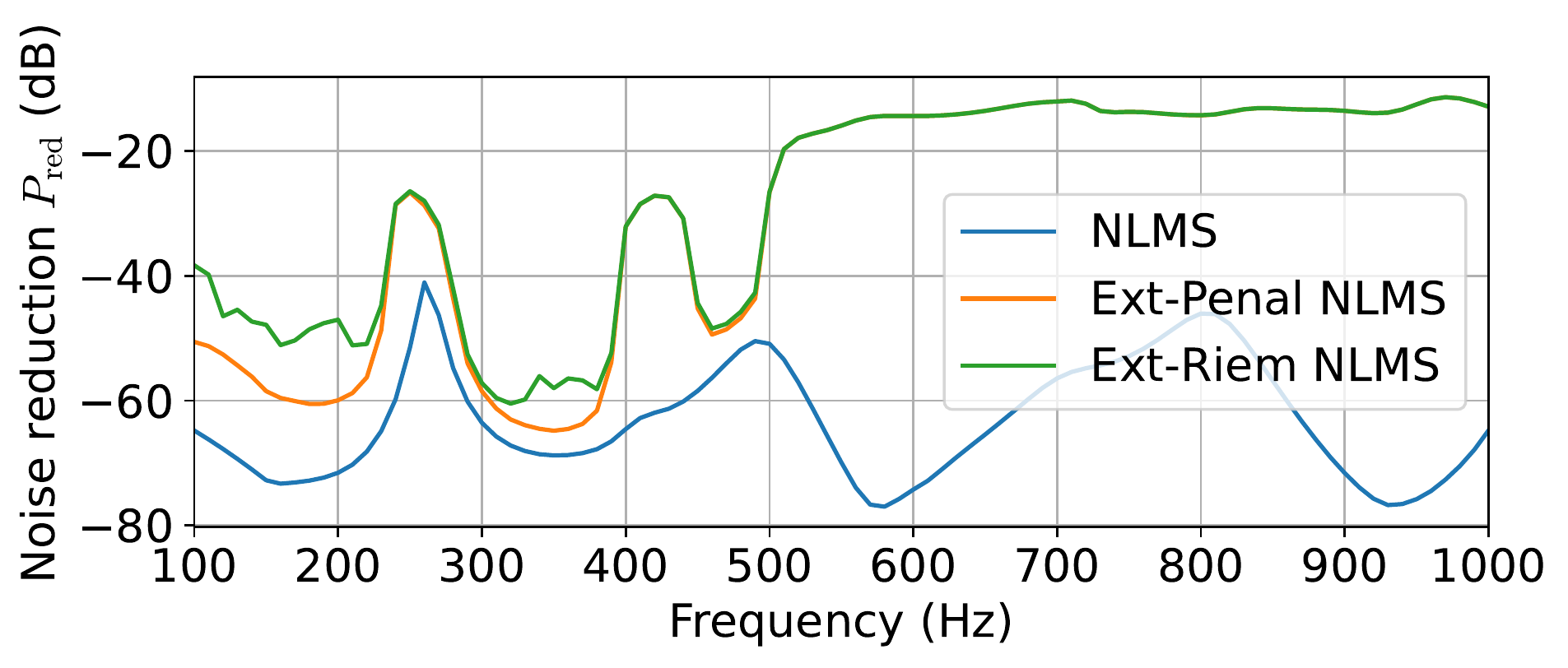}}
    \centerline{(a) Noise reduction $P_{\mathrm{red}}$}
  \end{minipage}
  \hfill
  \begin{minipage}[b]{\linewidth}
    \centering
    \centerline{\includegraphics[width=8.4cm]{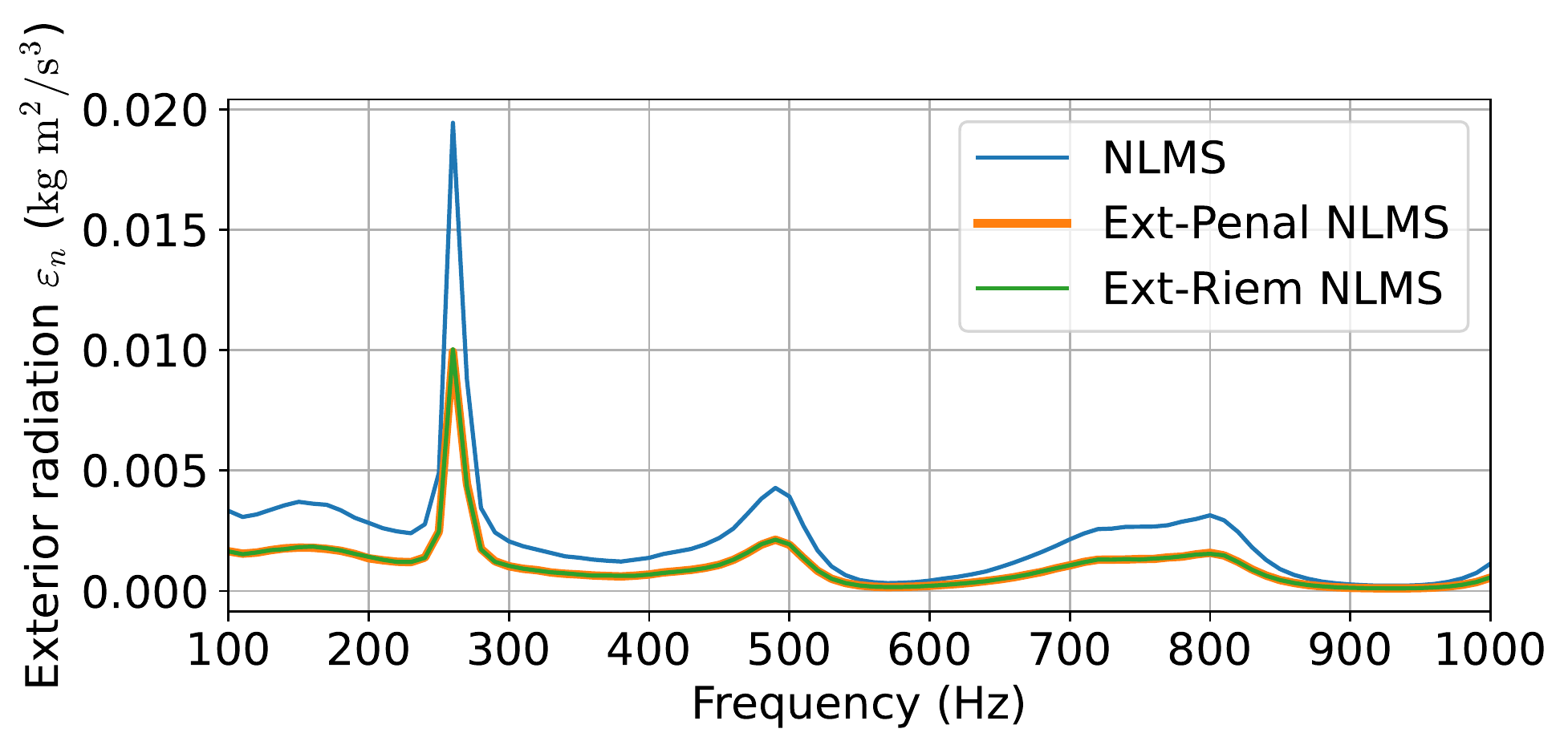}}
    \centerline{(b) Exterior radiation $\varepsilon_n$}
  \end{minipage}
  \caption{Noise reduction $P_{\mathrm{red}}$ and exterior radiation $\varepsilon_n$ after 50000 iterations with respect to frequency.}
  \label{fig:freq-performance}
\end{figure}

\begin{figure}[t]
  \begin{minipage}[b]{1.0\linewidth}
    \centering
    \centerline{\includegraphics[width=8.4cm]{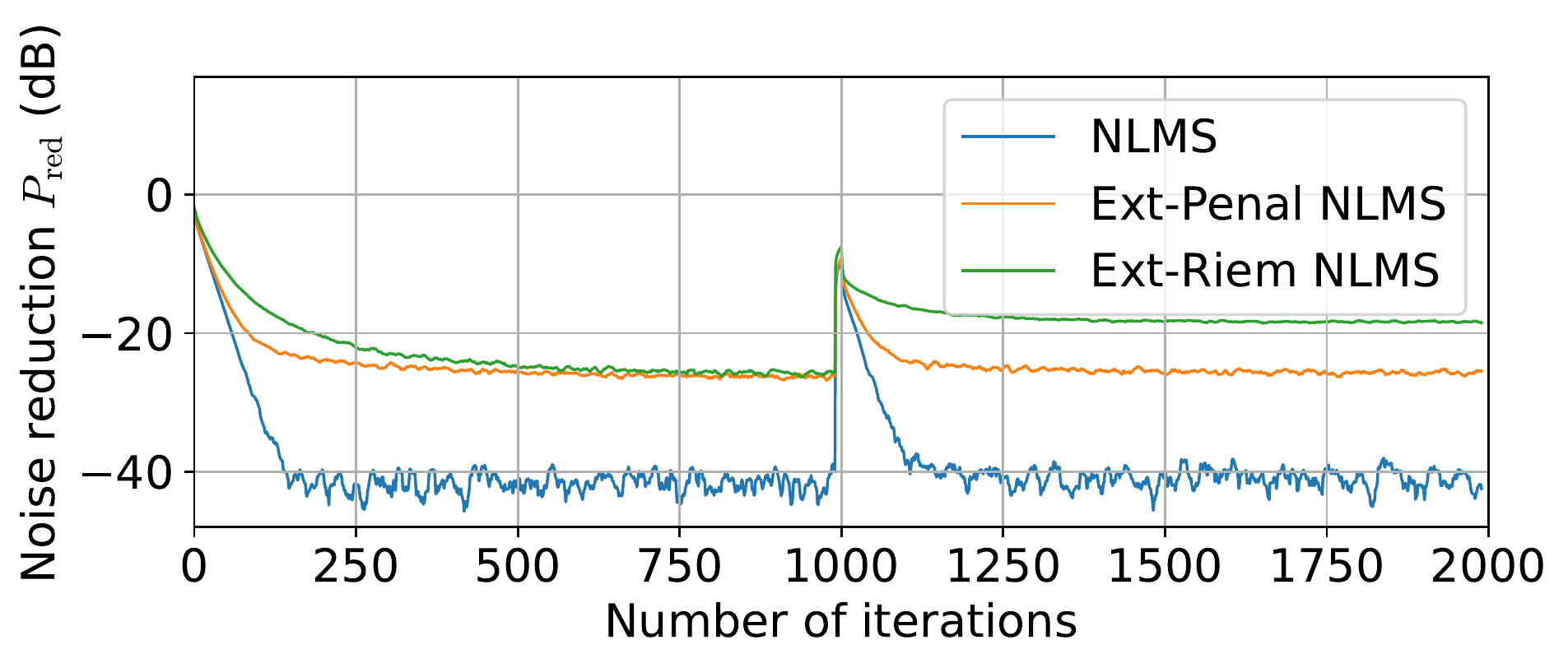}}
    \centerline{(a) Noise reduction $P_{\mathrm{red}}$}
  \end{minipage}
  \begin{minipage}[b]{1.0\linewidth}
    \centering
    \centerline{\includegraphics[width=8.4cm]{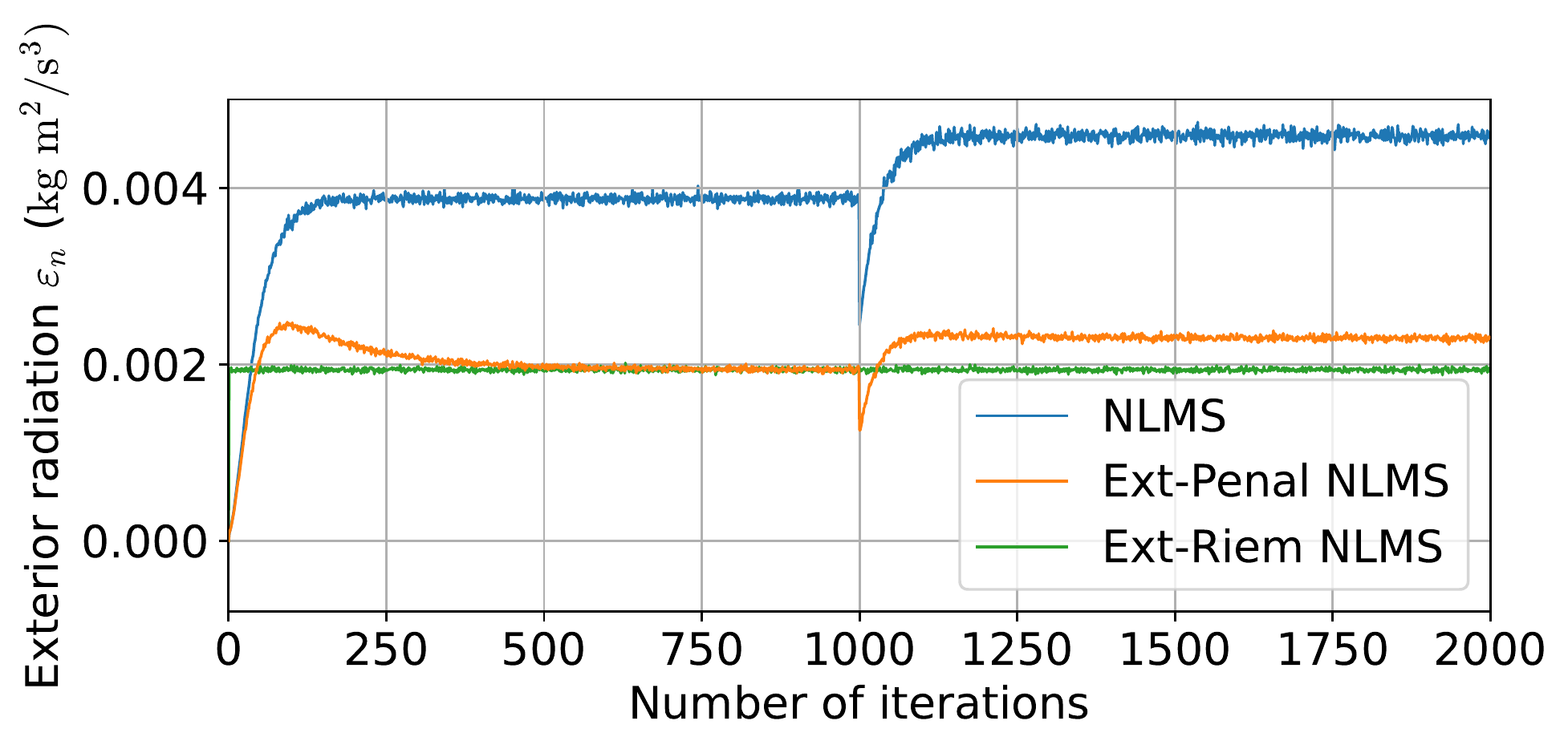}}
    \centerline{(b) Exterior radiation $\varepsilon_n$}
  \end{minipage}
  \caption{Noise reduction $P_{\mathrm{red}}$ and exterior radiation $\varepsilon_n$ at each iteration ($500~\mathrm{Hz}$) when the amplitudes of the primary noise sources changed after 25000 iterations.}
  \label{fig:result-500Hz-switch}
\end{figure}

\subsection{Settings}
\label{subsec:setting}

We assumed that two primary noise sources were placed at $(-3.0, 0.5)~\mathrm{m}$ and $(3.0, 0.0)~\mathrm{m}$. As shown in Fig.~\ref{fig:setting}, $M=4$ error microphones were set at $(\pm 0.5,~\pm 0.5)~\mathrm{m}$. $L=12$ secondary loudspeakers were regularly placed at the boundaries of two circular regions with a radius of $1.0~\mathrm{m}$ and $1.2~\mathrm{m}$. $R=2$ reference microphones were assumed to directly and separately obtain the primary noise signals. The primary and secondary sources were assumed to be point sources. The sound speed and medium density were set as $c = 343~\mathrm{m/s}$ and $\rho=1.3~\mathrm{kg/m^3}$, respectively. We also added Gaussian noise of $40~\mathrm{dB}$ $\mathrm{SNR}$ to the reference and error signals at each time frame.

The parameter $\gamma$ in $\sigma_n$ was set as the value of $10^{-4}$ times the maximum eigenvalue of $\bm{G}^\H\bm{G}$. The normalized step size parameter $\mu_0$ was set to $1.0$ for all the methods. The parameters $\lambda$ in \eqref{eq:Jpenalty} and $C$ in \eqref{eq:MPC-equality-problem} used for the constraint on the exterior radiation power were determined so that the exterior radiation power corresponds to half that obtained by the NLMS algorithm without exterior radiation suppression, i.e., \textbf{NLMS}, after convergence. Note that the setting of $\lambda$ requires exhaustive search as opposed to the setting of $C$, which can be simply determined from the exterior radiation power of \textbf{NLMS}.

In \textbf{Ext-Riem NLMS}, the noise at the error microphones can be amplified  particularly at the beginning of the adaptation process, because the control filter can be at a point on $\mathcal{M}$. Therefore, $\bm{d}_n$ and $\bm{e}_n$ are predicted from $\bm{x}_n$ before updating $\bm{W}_n$, and we set $\bm{y}_n=\bm{0}$ when $\|\bm{e}_n\|_2^2>\|\bm{d}_n\|_2^2$ is inferred.



As an evaluation measure for the noise reduction at the error microphone positions, we define $P_{\mathrm{red}}$ as
\begin{align}
  P_\mathrm{red}=\frac{\|\bm{e}_n\|_2^2}{\|\bm{d}_n\|_2^2}=\frac{\|\bm{d}_n+\bm{G}\bm{W}_n\bm{x}_n\|_2^2}{\|\bm{d}_n\|_2^2}.
\end{align}
The exterior radiation suppression is evaluated using $\varepsilon_n$ defined in \eqref{eq:Jext-2}. Note that $\varepsilon_n$ accurately evaluates the exterior radiation power in this setting since the secondary loudspeakers were assumed to be point sources. 



\subsection{Results}
\label{subsec:Results}

Fig.~\ref{fig:result-500Hz-normal} shows the moving averaged $P_{\mathrm{red}}$ and $\varepsilon_n$ at each iteration when the amplitudes of primary sources were $10.0$ and $5.0$, respectively, and the frequency was $500~\mathrm{Hz}$. As shown in Fig.~\ref{fig:result-500Hz-normal}(a), $P_{\mathrm{red}}$ was successfully reduced in the three methods, and its convergence speed was almost the same, although $P_{\mathrm{red}}$ of \textbf{Ext-Penal NLMS} and \textbf{Ext-Riem NLMS} were higher than that of \textbf{NLMS}. Meanwhile, the exterior radiation power $\varepsilon_n$ values of \textbf{Ext-Penal NLMS} and \textbf{Ext-Riem NLMS} after convergence were almost half that of \textbf{NLMS} as intended (Fig.~\ref{fig:result-500Hz-normal}(b)). In particlar, $\varepsilon_n$ was constant at the target value in \textbf{Ext-Riem NLMS} from the beginning of the ANC process, excluding the effect of sensor noise. 




The evaluation measures after convergence, which were averaged over 100 iterations, with respect to frequency are plotted in Fig.~\ref{fig:freq-performance}. The parameters $\lambda$ in \eqref{eq:Jpenalty} and $C$ in \eqref{eq:MPC-equality-problem} were determined at each frequency. The noise reduction performance characteristics of the \textbf{Ext-Penal NLMS} and \textbf{Ext-Riem NLMS} were slightly lower than that of NLMS at low frequencies, but their difference was increased at high frequencies. The exterior radiation powers of \textbf{Ext-Penal NLMS} and \textbf{Ext-Riem NLMS} were almost half that of \textbf{NLMS} for all the frequencies. Note that the exterior radiation power during the adaptation process was explicitly constrained only in \textbf{Ext-Riem NLMS}. 


Next, we show the result obtained when the amplitude of the primary sources was changed from $10.0$ and $5.0$ to $5.0$ and $10.0$ at $n=1000$ in Fig.~\ref{fig:result-500Hz-switch}. Note that $\|\bm{x}_n\|_2^2$ remained the same after the amplitude change. $P_{\mathrm{red}}$ of \textbf{NLMS} was smaller than those of \textbf{Ext-Penal NLMS} and \textbf{Ext-Riem NLMS} at $n=2000$. The exterior radiation power $\varepsilon_n$ was amplified in \textbf{NLMS} and \textbf{Ext-Penal NLMS} after $n=1000$. In contrast, $\varepsilon_n$ of \textbf{Ext-Riem NLMS} remained the same after the amplitude change, which can be considered as an advantage of the Riemannian-optimization-based algorithm. 


\section{Conclusion}
\label{sec:conclusion}

We proposed a multichannel ANC method for suppressing the exterior radiation of secondary loudspeakers while reducing noise at positions of error microphones. By using a representation of the exterior radiation power by a quadratic of  loudspeaker driving signals, we derived an NLMS algorithm based on the Riemannian optimization to update the control filter with the exterior radiation power constrained to a target value. The benefits of the proposed method against the method using a penalty term for the exterior radiation are as follows. 1) The exterior radiation power can be constrained to a target value during the adaptation process. 2) The parameter for the constraint can be set in advance (the exhaustive search for the balancing parameter in the penalty-term-based method is unnecessary). They are also shown in the experimental results. The formulation of the adaptive filtering algorithm for broadband ANC will be a future work. 


\section{Acknowledgment}
This work was supported by JST FOREST Program (Grant Number JPMJFR216M, Japan),  JSPS KAKENHI Grant Number JP22H03608, and Tateisi Science and Technology Foundation.


\bibliographystyle{IEEEtran}
\bibliography{str_def_abrv,koyama_en,ref_add}

\begin{thebibliography}{10}
\providecommand{\url}[1]{#1}
\csname url@samestyle\endcsname
\providecommand{\newblock}{\relax}
\providecommand{\bibinfo}[2]{#2}
\providecommand{\BIBentrySTDinterwordspacing}{\spaceskip=0pt\relax}
\providecommand{\BIBentryALTinterwordstretchfactor}{4}
\providecommand{\BIBentryALTinterwordspacing}{\spaceskip=\fontdimen2\font plus
\BIBentryALTinterwordstretchfactor\fontdimen3\font minus
  \fontdimen4\font\relax}
\providecommand{\BIBforeignlanguage}[2]{{%
\expandafter\ifx\csname l@#1\endcsname\relax
\typeout{** WARNING: IEEEtran.bst: No hyphenation pattern has been}%
\typeout{** loaded for the language `#1'. Using the pattern for}%
\typeout{** the default language instead.}%
\else
\language=\csname l@#1\endcsname
\fi
#2}}
\providecommand{\BIBdecl}{\relax}
\BIBdecl

\bibitem{nelson1991active}
P.~A. Nelson and S.~J. Elliott, \emph{Active control of sound}.\hskip 1em plus
  0.5em minus 0.4em\relax London: Academic Press, 1991.

\bibitem{kuo1999active}
S.~M. Kuo and D.~R. Morgan, ``Active noise control: a tutorial review,''
  \emph{Proc. {IEEE}}, vol.~87, no.~6, pp. 943--973, 1999.

\bibitem{kajikawa_gan_kuo_2012}
Y.~Kajikawa, W.-S. Gan, and S.~M. Kuo, ``Recent advances on active noise
  control: open issues and innovative applications,'' \emph{APSIPA Trans.
  Signal Inf. Process.}, vol.~1, p.~e3, 2012.

\bibitem{Samarasinghe_2016_InsideCabin}
P.~N. Samarasinghe, W.~Zhang, and T.~D. Abhayapala, ``Recent advances in active
  noise control inside automobile cabins: Toward quieter cars,'' \emph{{IEEE}
  Signal Process. Mag.}, vol.~33, no.~6, pp. 61--73, 2016.

\bibitem{Zhang_2018_ANCoverSpace}
J.~Zhang, T.~D. Abhayapala, W.~Zhang, P.~N. Samarasinghe, and S.~Jiang,
  ``Active noise control over space: A wave domain approach,'' \emph{{IEEE/ACM}
  Trans. Audio, Speech, Lang. Process.}, vol.~26, no.~4, pp. 774--786, 2018.

\bibitem{ma2020}
F.~Ma, W.~Zhang, and T.~D. Abhayapala, ``Active control of outgoing broadband
  noise fields in rooms,'' \emph{{IEEE/ACM} Trans. Audio, Speech, Lang.
  Process.}, vol.~28, pp. 529--539, 2020.

\bibitem{Koyama:IEEE_ACM_J_ASLP2021}
S.~Koyama, J.~Brunnstr\"{o}m, H.~Ito, N.~Ueno, and H.~Saruwatari, ``Spatial
  active noise control based on kernel interpolation of sound field,''
  \emph{{IEEE/ACM} Trans. Audio, Speech, Lang. Process.}, vol.~29, pp.
  3052--3063, 2021.

\bibitem{rafaely2000computationally}
B.~Rafaely and S.~J. Elliot, ``A computationally efficient frequency-domain
  {LMS} algorithm with constraints on the adaptive filter,'' \emph{{IEEE}
  Trans. Signal Process.}, vol.~48, no.~6, pp. 1649--1655, 2000.

\bibitem{qiu2001study}
X.~Qiu and C.~H. Hansen, ``A study of time-domain {FXLMS} algorithms with
  control output constraint,'' \emph{J. Acoust. Soc. Amer.}, vol. 109, no.~6,
  pp. 2815--2823, 2001.

\bibitem{shi2019two}
D.~Shi, W.-S. Gan, B.~Lam, and C.~Shi, ``Two-gradient direction {FXLMS}: An
  adaptive active noise control algorithm with output constraint,'' \emph{Mech.
  Syst. Signal Process.}, vol. 116, pp. 651--667, 2019.

\bibitem{shi2021opt}
D.~Shi, W.-S. Gan, B.~Lam, S.~Wen, and X.~Shen, ``Optimal output-constrained
  active noise control based on inverse adaptive modeling leak factor
  estimate,'' \emph{{IEEE/ACM} Trans. Audio, Speech, Lang. Process.}, vol.~29,
  pp. 1256--1269, 2021.

\bibitem{Ueno:ICASSP2018}
N.~Ueno, S.~Koyama, and H.~Saruwatari, ``Sound field reproduction with exterior
  radiation cancellation using analytical weighting of harmonic coefficients,''
  in \emph{Proc. {IEEE} Int. Conf. Acoust., Speech, Signal Process.
  ({ICASSP})}, 2018, pp. 466--470.

\bibitem{Arikawa:ICA2022}
K.~Arikawa, S.~Koyama, and H.~Saruwatari, ``Kernel-interpolation-based spatial
  active noise control with exterior radiation suppression,'' in \emph{Proc.
  Int. Congr. Acoust. ({ICA})}, 2022.

\bibitem{AbsMahSep2008}
P.-A. Absil, R.~Mahony, and R.~Sepulchre, \emph{Optimization Algorithms on
  Matrix Manifolds}.\hskip 1em plus 0.5em minus 0.4em\relax Princeton:
  Princeton University Press, 2008.

\bibitem{sato2021riemannian}
H.~Sato, \emph{Riemannian Optimization and Its Applications}.\hskip 1em plus
  0.5em minus 0.4em\relax Cham: Springer Nature Switzerland AG, 2021.

\bibitem{stochastic}
S.~Bonnabel, ``Stochastic gradient descent on {Riemannian} manifolds,''
  \emph{{IEEE} Trans. Autom. Control}, vol.~58, no.~9, pp. 2217--2229, 2013.

\bibitem{haykin2013adaptive}
S.~Haykin, \emph{Adaptive Filter Theory: International Edition, 5/E}.\hskip 1em
  plus 0.5em minus 0.4em\relax London: Pearson, 2013.

\bibitem{Shustin2019_projection}
B.~Shustin and H.~Avron, ``Preconditioned {R}iemannian optimization on the
  generalized stiefel manifold,'' 2019.

\bibitem{horn2012matrix}
R.~A. Horn and C.~R. Johnson, \emph{Matrix analysis}.\hskip 1em plus 0.5em
  minus 0.4em\relax New York: Cambridge University Press, 2012.

\bibitem{Sato2019}
\BIBentryALTinterwordspacing
H.~Sato and K.~Aihara, ``Cholesky {QR}-based retraction on the generalized
  {S}tiefel manifold,'' \emph{Comput. Optim. App.}, vol.~72, no.~2, pp.
  293--308, 2019. [Online]. Available:
  \url{https://doi.org/10.1007/s10589-018-0046-7}
\BIBentrySTDinterwordspacing

\bibitem{bartels1972solution}
R.~H. Bartels and G.~W. Stewart, ``Solution of the matrix equation {$AX+ XB=
  C$},'' \emph{Commun. ACM}, vol.~15, no.~9, pp. 820--826, 1972.

\end{thebibliography}

\end{document}